\documentclass[%
superscriptaddress,
preprint,
amsmath,amssymb,
prb,
]{revtex4-2}

\usepackage{graphicx}
\usepackage{bm}
\usepackage{bbm}
\usepackage[utf8]{inputenc}
\usepackage{hyperref}
\usepackage[a4paper,margin=20mm]{geometry}
\usepackage{adjustbox}
\usepackage{siunitx}

\date{\today}

\begin{document}

\def\mytitle{Ultrafast dynamics of moments in bulk ferromagnets}
\def\myfirstauthor{Mouad Fattouhi}
\def\mysecondauthor{Pascal Thibaudeau}
\def\mythirdauthor{Liliana D. Buda-Prejbeanu}
\def\spintec{Univ. Grenoble Alpes, CEA, CNRS, Grenoble-INP, SPINTEC, 38000 Grenoble, France}
\def\dam{CEA, DAM, Le Ripault, F-37260, Monts, France}

\title{\mytitle}
\author{\myfirstauthor}
\email{mouad.fattouhi@cea.fr}
\affiliation{\spintec}
\author{\mysecondauthor}
\affiliation{\dam}
\author{\mythirdauthor}
\affiliation{\spintec}

\begin{abstract} 
    A robust and efficient model for investigating the ultrafast dynamics of magnetic materials excited by laser pulses has been created, integrating dynamic Landau-Lifshitz-Bloch equations with a quantum thermostat and a two-temperature model. The model has been successfully applied to three archetypal materials in the literature: nickel, cobalt, and iron. Additionally, analysis of the ultrafast dynamic susceptibility tensor indicates that off-diagonal components display specific features depending on whether a continuous external magnetic field is present. 
\end{abstract}

\maketitle

%%%%%%%%%%%%%%%%%%%%%%%%%%%%%%%%%%%%%%%%%%%%%%%%%%%%%%%%%%%%%%%%%
\section{Introduction}\label{sec:introduction}
%%%%%%%%%%%%%%%%%%%%%%%%%%%%%%%%%%%%%%%%%%%%%%%%%%%%%%%%%%%%%%%%%

Nowadays, the precise control of thermal fluctuations of relevant quantities to operate reliably small magnetic devices remains challenging ~\cite{barlaSpintronicDevicesPromising2021,dienyOpportunitiesChallengesSpintronics2020,johnMagnetisationSwitchingFePt2017}.
Describing these fluctuations accurately is also a pressing issue, with a long history of computational methods developed to address it~\cite{neelThermoremanentMagnetizationFine1953,brownThermalFluctuationsSingleDomain1963,coffeyThermalFluctuationsMagnetic2012}.
The standard approach for their description is stochastic calculus~\cite{vankampenStochasticDifferentialEquations1976,gardinerStochasticMethodsHandbook2009}: the quantities of interest depend on the correlation functions of the magnetization. 
These correlations are primarily obtained today either from numerical simulations~\cite{chenSpindynamicsStudyDynamic1994,simonSpincorrelationsMagneticStructure2014} or from the acquisition and processing of experimental data~\cite{fristonMovementRelatedEffectsFMRI1996,camsariChargeSpinSpin2020,safranskiDemonstrationNanosecondOperation2021}.

The ferromagnetic phase of simple transition metals is one of the popular benchmark systems serving to explore abilities of various theories and models of magnetism. 
For example, the relaxation processes of electrons and spins systems following the absorption of femtosecond optical pulses in simple ferromagnetic metals have been studied using optical and magneto-optical pump-probe techniques~\cite{vomirRealSpaceTrajectory2005,unikandanunniAnisotropicUltrafastSpin2021}, leading to a growing discipline called femtomagnetism~\cite{kirilyukUltrafastOpticalManipulation2010,zhangAllopticalSpinSwitching2016}.
These experiments trigger complex out-of-equilibrium phenomena when a pulsed-laser excitation scheme leads to the swift creation of spin currents, entering the realm of ultrafast spintronics~\cite{malinowskiControlSpeedEfficiency2008,melnikovUltrafastTransportLaserExcited2011,rudolfUltrafastMagnetizationEnhancement2012}. 
This advancement paves the way for THz spintronic devices, marking the first practical use of this ultrafast spin physics to manipulate ultrafast magnetism~\cite{walowskiPerspectiveUltrafastMagnetism2016}.

Optimization and reliable control of ultrafast magnetic devices is only possible by theoretical understanding.
Unfortunately, due to the complexity of the processes involved at different length and time scales, no single model of ultrafast magnetization dynamics exists to date.
Based on the physical contents, theoretical models can generally be classified into three main categories: (i) the direct coherent interaction between the laser pulse and the spin degree of freedom~\cite{zhangLaserInducedUltrafastDemagnetization2000,crouseillesVlasovMaxwellEquations2023}, (ii) local spin dynamics as triggered by laser heating or excitation~\cite{beaurepaireUltrafastSpinDynamics1996,koopmansExplainingParadoxicalDiversity2010,manchonTheoryLaserinducedDemagnetization2012,nievesQuantumLandauLifshitzBlochEquation2014} and (iii) non-local transfer of angular momentum such as superdiffusive spin transport~\cite{battiatoSuperdiffusiveSpinTransport2010}.

The direct processes from category (i) are assumed to play minor roles with the naive argument that the dominant magnetization dynamics happens on a timescale larger than the typical femtosecond duration of the laser pulse. 
To support this argument, it has been shown experimentally that the demagnetization dynamics is not affected by the laser pump helicity in such short durations~\cite{dallalongaInfluencePhotonAngular2007}.
In contrast, categories (ii) and (iii) include the effect of out-of-equilibrium electronic distributions generated by the laser as projected descriptions of the dynamics of atomistic magnetic moments.
Since the time scales of demagnetization experiments are fluence-dependent, when strong demagnetization is reached beyond half the saturation magnetization, it was necessary to include fluctuations in atomistic magnetic moments.
Additionally, simulations of an excited spin system using thermal noise that mimics fluctuations have shown the significance of high-energy THz spin waves in remagnetization processes~\cite{zhangUltrafastTerahertzMagnetometry2020,rongioneEmissionCoherentTHz2023}.

Most thermal spin models use direct simulation of atomic spins, either in a stochastic~\cite{erikssonAtomisticSpinDynamics2017} or deterministic~\cite{thibaudeauThermostattingAtomicSpin2011} form.
Because the value of magnetic moments are related to the electronic properties of solids, the thermal connection associates the electronic temperature with the spin system via an electron-spin relaxation rate.
Microscopically, exciting the electron system yields an increased rate of various scattering processes that may involve spin flips, such as electron-phonon scattering~\cite{giustinoElectronphononInteractionsFirst2017}, electron-magnon scattering~\cite{raquetElectronmagnonScatteringMagnetic2002,haagRoleElectronmagnonScatterings2014}, and electron-electron scattering~\cite{fahnleElectronTheoryFast2011}.
The consensus is that spin-flip scattering processes contribute to a single effective electron-spin relaxation rate, that enters in various forms of time-dependent evolutions of heat reservoirs~\cite{anisimovElectronEmissionMetal1974,beaurepaireUltrafastSpinDynamics1996}.
However, the way the temperature of the electron system is coupled to either the stochastic or deterministic form of the atomistic spin dynamics is still an issue~\cite{chubykaloLangevinDynamicSimulation2002}. 

For the atomistic spin system, since the Hamiltonian operator is known, both atomic Landé factors and magnetic coupling parameters, including exchange energy and magnetic anisotropy, can be derived from {\it ab initio} calculations.
It is well established that for a spin ensemble connected to heat reservoirs, the magnitude of the magnetization is not conserved, and longitudinal damping effects must be considered~\cite{kazantsevaMultiscaleModelingMagnetic2008,mcdanielApplicationLandauLifshitzBlochDynamics2012,voglerLandauLifshitzBlochEquationExchangecoupled2014}.
Consequently, the deterministic thermal model developed initially was a thermal macrospin model~\cite{garaninThermalFluctuationsLongitudinal2004}, that transforms the dynamical equations for an ensemble of interacting spins to an equation of motion for a single macrospin where, in order to account for the thermal dynamics, an additional Bloch-like relaxation term is found.
An equation describing the precession of the magnetization dynamics of nanomagnets and accounting for both transverse and longitudinal forms of damping is commonly referred to as a Landau-Lifshitz-Bloch (LLB) equation~\cite{garaninDynamicsEnsembleSingledomain1990,garaninFokkerPlanckLandauLifshitzBlochEquations1997}.
This equation has shown an increasing relevance in magnetism because of its capability to describe several classes of out-of-equilibrium phenomena where thermal excitation is essential~\cite{atxitiaFundamentalsApplicationsLandau2017}.
The primary challenge with the LLB equation is not the absence of microscopic material parameters, but rather the lack of effective thermodynamic functions that are derived from these microscopic parameters.
Prior to any simulations involving the LLB equation, it is essential to provide these functions, particularly the temperature dependence of the exchange stiffness, spontaneous magnetization, and longitudinal and transverse damping parameters, below and above the Curie Temperature for ferromagnets.

Tranchida {\it et al.} ~\cite{tranchidaHierarchiesLandauLifshitzBlochEquations2018} have formulated a rigorous and systematic statistical averaging procedure, based on functional calculus and incorporating a few microscopic material parameters.
This procedure can be applied to derive set of LLB-like differential equations for the first and second moments of spin variables.
The differential equation governing the dynamics of the first moment preserves the original form of the LLB equation and inherently generates a thermal longitudinal damping. 
Damping is calculated on the basis of variations in magnetic moments due to thermal noise, thus avoiding the need to provide prior information on longitudinal and transverse thermal damping.
This equation also includes a second moment that does not factorize in a sum of first moments, suggesting that the magnetization variance may also fluctuate over time.
In this model called dynamical LLB (dLLB), the damping torque that influences the precession of the magnetization, is governed by the dynamics of both the first and second moments.
In return, the second moment also fluctuates in time and is itself the solution of a differential equation.
The entire differential system constitutes an open hierarchy that can be closed in various ways~\cite{tranchidaHierarchiesLandauLifshitzBlochEquations2018} and solved numerically in a consistent manner.

Once the fluctuation-dissipation theorem, that establishes a relationship between the spontaneous fluctuations of a magnetic system and its response to thermally driven external perturbations~\cite{kuboFluctuationdissipationTheorem1966}, is satisfied, it has been shown that the solution of the dLLB equations allows the computation to any thermal expectation value of the magnetization consistently. 
Consequently, this strategy makes it easier to predict the temperature dependence of exchange stiffness, susceptibilities, spontaneous magnetization and damping parameters, at equilibrium or for any non-stationary process.

In summary, this paper presents the development of a consistent dLLB model that serves as a platform for reliable predictions of ultrafast demagnetization processes in magnetic metals.

%%%%%%%%%%%%%%%%%%%%%%%%%%%%%%%%%%%%%%%%%%%%%%%%%%%%%%%%%%%%%%%%%%
\section{Theoretical framework}\label{sec:theory}
%%%%%%%%%%%%%%%%%%%%%%%%%%%%%%%%%%%%%%%%%%%%%%%%%%%%%%%%%%%%%%%%%%

To accurately simulate the ultrafast magnetization dynamics caused by laser absorption, an essential component in atomistic simulations of magnetization dynamics is needed~\cite{antropovSpinDynamicsMagnets1996}.
The so-called phenomenological three-temperature model (3TM)~\cite{beaurepaireUltrafastSpinDynamics1996} considers three interacting reservoirs, namely hot electrons, phonons and spins, and their corresponding temperatures, that exchange heat over time~\cite{pankratovaHeatconservingThreetemperatureModel2022}.
{When a laser interacts with a metal, it generates electromagnetic waves that heat the metal. 
The laser energy is primarily absorbed by free electrons, leading to a rapid thermalization process at fs scale. 
Heavy ions do not directly absorb the laser energy, but the heated electrons transfer a small fraction of their energy to the lattice, leading to lattice heating.
Then, the spins and the phonons become thermally unbalanced, and their temperatures change at a rate defined by their heat capacities and the strength of their coupling to the electron reservoir, as well as interactions among themselves.
}

Spinning particles produce entropy that induces a microcanonical temperature of this system~\cite{nurdinDynamicalTemperatureSpin2000}. 
According to the fluctuation-dissipation theorem, this spin temperature is also related to the noise amplitude, often labelled $D$, of the thermal field that enters the stochastic precession equation.
Consequently, the thermal connection to the dLLB system, which mimics magnetic behavior, can be safely established.
Both the 3TM and dLLB differential equations can be integrated simultaneously in a coupled manner, so we can set the noise amplitude $D$ to be a function of the temperature of the spins computed by the 3TM dynamical system. 
However, following this scheme, the dLLB model cannot provide feedback on the effective physical quantities that appear in the 3TM.

For each atomic site $i$, the time evolution of magnetic moments is obtained solving dLLB closed set of differential equations, as described in Ref.~\cite{tranchidaHierarchiesLandauLifshitzBlochEquations2018}.
To simplify the reading and the manipulation of the expressions, we introduce several quantities of interest. 
First we set ${\bm s}_i\equiv\langle{\bm S}_i\rangle$ to be the average-over-the-noise magnetic moment direction ${\bm S}_i$. 
In general, the local precession vector ${\bm\omega_i}$ is a random variable since it is a functional of all the random magnetic moments ${\bm S}_j$.
The complexity increases in the behavior of all functionals that include effective fields and magnetic moments~\cite{tranchidaClosingHierarchyNonMarkovian2016,thibaudeauNonMarkovianMagnetizationDynamics2016}. 
In situations where the local precession vector remains close to its average value, its expression can be simplified by assuming that it is not a random variable, but rather a function of the average-over-noise moments direction only.
This is called the mean-field approximation (MFA), which involves both microscopic exchange parameters $J_{ij}$ of the material considered, and the first moment ${\bm s}_j$ for any neighboring sites $j$.
In equation this reads 
{
    \begin{align}
        \bm{\omega}_i&\equiv\frac{1}{\hbar}\sum_{j\neq i}J_{ij}{\bm s}_j.
        \label{eq:MFA}
    \end{align}
}

This allows the formation of a skew-symmetric matrix $\Omega$ where $\Omega_{IJ}\equiv\epsilon_{IJK}\omega_i^K$ are its components {with $\epsilon_{IJK}$ being the components of the Levi-Civita antisymmetric tensor with $I,J,K\equiv x,y,z$ and $i$ being the atomic site.}
We also set $\Sigma_i\equiv\langle{\bm S}_i\otimes{\bm S}_i\rangle$ a symmetric $3\times 3$ matrix, a symmetric matrix $\Gamma_i\equiv{\bm s}_i\otimes \bm{s}_i$ and a (general) matrix $M_i\equiv{\bm s}_i\otimes{\bm \omega}_i$, which can be decomposed into a symmetric (resp. antisymmetric) part labelled ${M}_i^S$ (resp.${M}_i^A$).
The dLLB model is reduced to the following set of equations

\begin{align}
    \left(1+\alpha^2\right)\frac{d{\bm s}_i}{dt}=&{\bm\omega}_i\times{\bm s}_i-\alpha\left({\Sigma}_i-{{\sf{Tr}}({\Sigma}_i)}{\mathbbm{1}}\right){\bm{\omega}_i}\nonumber\\
    &-\frac{2D}{(1+\alpha^2)}{\bm s}_i\label{dLLB1}\\
    \left(1+\alpha^2\right)\frac{d{{\Sigma}_i}}{dt}&=2{{\Omega}}\,{\Sigma}_i\nonumber\\
    &-2\alpha\left([{M}_i^A,{\Sigma_i}]+{\sf{Tr}}({M}_i)({\Sigma}_i-2{\Gamma}_i)\nonumber\right.\\
    &\left.-({\sf{Tr}}({\Sigma}_i)-2{\sf{Tr}}({\Gamma}_i)){M}_i^S\right)
    \nonumber\\
    &-\frac{2D}{(1+\alpha^2)}\left(3{{\Sigma}_i}-{\mathbbm{1}}{{\sf{Tr}}({\Sigma}_i)}\right)\label{dLLB2}
\end{align}
where $\alpha$ is the dimensionless transverse Gilbert damping parameter and ${\sf{Tr}}$ stands for the trace of the matrix.
In eq.~\eqref{dLLB2}, the commutator of two matrices is appearing as an operator that produces a matrix by contracting its internal index, thereby reducing it to a symmetric matrix, as expected.
Equations~\eqref{dLLB1} and \eqref{dLLB2} are numerically integrated over time by a global Runge-Kutta scheme of 4-th order, with an either constant or varying timestep. {The dLLB equations are a result of statistical averaging over the noise of the well known sLLG equation. The details about the statistical averaging procedure and the model derivation are presented in our previous works \cite{tranchidaClosingHierarchyNonMarkovian2016,thibaudeauNonMarkovianMagnetizationDynamics2016,tranchidaHierarchiesLandauLifshitzBlochEquations2018}.}

{Physically, Eq.~\eqref{dLLB1} describes the dynamics of ${\bm s}_i$, the mean of the stochastic magnetic moment vector $\bm{S}_i$. 
The initial term denotes the torque associated with the precession of the magnetic moment.
The subsequent term takes into account the transverse relaxation that involves a $3 \times 3$ matrix ${\Sigma}_i$ and the final term represents the longitudinal relaxation.
Equation \eqref{dLLB2} governs the dynamics of ${\Sigma}_i$, which represents the statistical variance of the stochastic magnetic moment $\bm{S}_i$. 
The first term describes a precession-like behavior, represented by matrix multiplication of a skew-symmetric matrix formed by the effective pulsation vector $\bm{\omega}_i$, and the matrix ${\Sigma}_i$. 
The second and third terms exhibit matrix transverse and longitudinal relaxation processes, respectively, similar to Equation \eqref{dLLB1}, except that is for the variance.
}  

To enlighten the physics of quantities entering Eqs.~\eqref{dLLB1} and~\eqref{dLLB2}, we note that in statistical physics, many extensive quantities—which are proportional to the volume or size of a given system—are related to cumulants of random variables~\cite{vankampenStochasticProcessesPhysics1992}.
As cumulants measure the strength of the interactions between random variables, they also measure their independence.
Thus, for each atomic site $i$, the first cumulant is the average-over-the-noise value of the moment and retains the same physical content than the first moment ${\bm s}_i$. 
The second cumulant, denoted as $\chi_i$ is a matrix that encodes the variance of the moment, thereby measuring the fluctuations of this moment induced by thermal noise.
In equation this reads,
\begin{align}
    \label{eq::second_moment}
    \chi_{i}(t)&\equiv\langle {\bm S}_i(t)\otimes{\bm S}_i(t)\rangle-\langle{\bm S}_i(t)\rangle\otimes\langle{\bm S}_i(t)\rangle\nonumber\\
    &=\Sigma_i(t)-\Gamma_i(t)
\end{align}
The volume average of this matrix can be computed as
\begin{equation}
\label{eq::Volume_second_moment}
 \chi(t)=\frac{1}{N}\sum_{i=1}^N\chi_i(t)\, .   
\end{equation}

%%%%%%%%%%%%%%%%%%%%%%%%%%%%%%%%%%%%%%%%%%%%%%%%%%%%%%%%%%%%%%%%%%
\section{Applications}\label{sec:applications}
%%%%%%%%%%%%%%%%%%%%%%%%%%%%%%%%%%%%%%%%%%%%%%%%%%%%%%%%%%%%%%%%%%

To achieve accurate predictions of thermal effects, we first benchmark the selected models in time when magnetic equilibrium is reached for any finite temperature.

\subsection{Magnetization curve of 3d ferromagnets}\label{sec:magnetization-curve}

The capacity of the dLLB model, as described by equations \eqref{dLLB1} and \eqref{dLLB2}, to precisely forecast the experimental transition temperature $(T_C)$ between a ferromagnetic state at low temperatures and a paramagnetic state at high temperatures is initially evaluated.
For this purpose, we simulate the Curie curve of some 3d ferromagnetic simple metals as fcc-Ni, bcc-Fe, and fcc-Co, by using magnetic parameters provided in Table~\ref{table:cp_elements}.
{To calculate $T_C$, we establish a real-space cell with $N$ spins in total with periodic boundary conditions so that the influence of each neighbor on every atomic site is taken into consideration mimicking a bulk compound.
The number of spins $N$ in the dLLB simulation should be sufficiently large to accurately account for all neighboring interactions. 
The specific size and number of atoms within each supercell are dependent on the unique characteristics of each material being investigated. For Ni and Co, we considered a supercell containing $5\times 5 \times 5$ unit cells along the (x, y, z)-axes, resulting in a simulated lattice with $N=500$ atomic spins.
Similarly, for Fe, we employed supercells consisting of $7\times 7\times 7$ unit cells, resulting in lattices with about $N=686$ atomic spins.}  
The equations \eqref{dLLB1} and \eqref{dLLB2} are integrated over time for each temperature $T$ within the range of 0 to 2000 Kelvin, continuing this process until the system achieves equilibrium at time $t_{eq}$.
The volume average of the first moment, labelled $\mathfrak{m}$, is evaluated according to

\begin{align}
  \mathfrak{m}&=\frac{1}{N}\sum_{i=1}^N\bm s_{i}(t_{eq}) 
  \label{eq::Volume_first_moment} 
\end{align}

\begin{figure}[htbp]
    \centering
    \includegraphics[height=.8\textheight,width=\columnwidth,keepaspectratio]{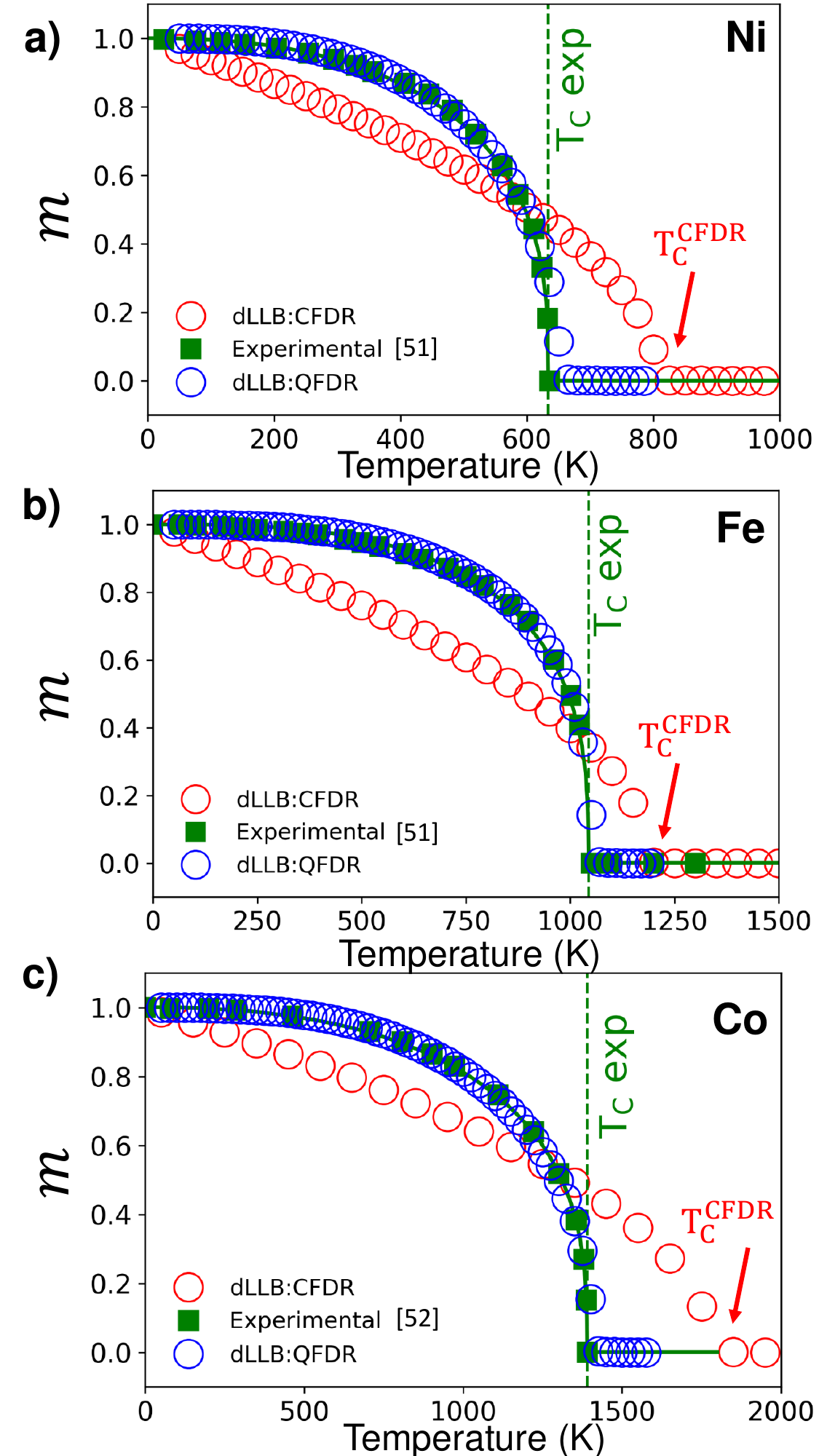}
    \caption{Norm of the volume average
    of the first moment as a function of an external temperature for a) Nickel, b) Iron, c) Cobalt, computed by atomistic spin dynamics of dLLB equations with both classical (CFDR) and quantum-corrected (QFDR) thermal baths. The experiments are taken from Ref.~\cite{crangleMagnetizationPureIron1971,kuzminShapeTemperatureDependence2005} (see text).}
    \label{figure:magnetization-curve}
\end{figure}

The results are presented in Fig.~\ref{figure:magnetization-curve} for Ni, Fe, and Co respectively.
In the figure, we plot the magnetic equilibrium state reached by the dLLB equations, when the external temperature $T$ is connected to the spins by either a classical fluctuation-dissipation relation (CFDR) or a quantum-like fluctuation-dissipation relation (QFDR).
The experimental measurements are taken from Ref.~\cite{crangleMagnetizationPureIron1971,kuzminShapeTemperatureDependence2005}.
For CFDR, the noise amplitude $D$, appearing in the dLLB equations, is related to the external temperature $T$ linearly~\cite{brownThermalFluctuationsSingleDomain1963} as 
\begin{align}
    D&=\frac{\alpha}{\hbar}k_BT
    \label{eq:CFDR}
\end{align}
\noindent where $\hbar$ is the reduced Planck constant and $k_B$ is the Boltzmann constant. 

Upon examination, it is shown that the CFDR (depicted as red circles) replicates a characteristic Langevin $M(T)$ curve. 
However, it yields an overestimated Curie temperature, denoted as $T_C^{\mathrm{CFDR}}$, which exceeds the experimental Curie temperature, $T_C^{\mathrm{exp}}$, for all three ferromagnets under investigation.
A thorough comparison with experimental data (represented by green squares) indicates that, besides overestimating the Curie temperature, the dLLB-CFDR model fails to accurately replicate the magnetization profile at lower temperatures, including the critical exponents near $T_c$.
The erroneous prediction of the Curie temperature can be ascribed to the MFA of exchange interactions~\cite{binneyTheoryCriticalPhenomena1992}. 
The fact of neglecting the spatial correlations between spins inhibits the excitation of magnons, resulting in inaccurate behavior in the vicinity of the critical temperature.
To address this issue, it is necessary to incorporate second-order fluctuations~\cite{georgesHowExpandMeanfield1991}, while MFA considers first-order fluctuations only~\cite{garaninSelfconsistentGaussianApproximation1996}. 
Previous studies have shown that accounting for second-order fluctuations can be simplified to a linear rescaling of the MFA exchange constant, expressed as $J_{ij} = \gamma_{SW} J_{ij}^{MFA}$, where $\gamma_{SW}$ represents the spin-wave mean-field correction. 
This correction has been calculated to be approximately $0.8$ for the 3d ferromagnetic elements \cite{garaninSelfconsistentGaussianApproximation1996,wysinOnsagerReactionfieldTheory2000}. 
The specific $\gamma_{SW}$ value of the ferromagnets used throughout our work are reported in Table~\ref{table:cp_elements}.  

It is not surprising that the dLLB-CFDR model produces an inaccurate low-temperature magnetization shape.
This limitation stems from a fundamental issue in classical atomistic spin dynamics related to the Langevin heat bath \cite{brownThermalFluctuationsSingleDomain1963}. 
The main issue with Langevin baths is that they use classical Maxwell-Boltzmann statistics, because of the linear dependence of the spin-bath coupling constant $D$ to the temperature $T$ as shown in Eq.~\eqref{eq:CFDR}, which allows all internal magnetic excitations to be equally populated.
Consequently, the energy spectrum of spin excitations is assumed to be continuous, allowing all possible spin orientations. 
In reality, the magnon energy spectrum is discrete at low temperatures due to quantization effects \cite{wooQuantumHeatBath2015}. 
To address this issue, we replace the Langevin bath described by equation ~\eqref{eq:CFDR} with a quantum heat bath. 
This alternative approach leverages Bose-Einstein statistics to better capture the underlying physics.
In this case, the expression for the spin-bath coupling depends on the magnon density of states (mDOS) $g_m(\omega,T)$.
In equation this reads
\begin{align}
    \label{eq::QL_heatbath}
    D&=\frac{\alpha}{\hbar}\int_{0}^{\infty}\frac{\hbar\omega}{\exp{\left(\frac{\hbar\omega}{k_\text{B}T}\right)-1}}g_m(\omega,T)\text{d}\omega
\end{align} 
where $\omega$ represents the magnon frequency. 
To apply the quantum heat bath, we need prior knowledge of a mDOS, or at least a representation of it, such as the one derived by Woo {\it et al.}~\cite{wooQuantumHeatBath2015}.  
Here, we consider a mDOS computed from a quartic magnon dispersion relation in k-space~\cite{zhitomirskyColloquiumSpontaneousMagnon2013} as 
\begin{align}
    \label{eq::mDOS_quartic}
    g_m(\omega,T)&=
    \begin{cases}
        \frac{\hbar V_\text{at}}{4\pi^{2}A\left(T\right)}\sqrt{\frac{1}{2\epsilon}\left(\sqrt{\frac{\omega_c}{\omega_c-\omega}}-1\right)} & \mathrm{for}\quad\omega\leqslant\omega_{c}\\
        0 & \mathrm{for}\quad\omega>\omega_{c}
    \end{cases}
\end{align}
where $V_\text{at}$ is the atomic cell volume, $\epsilon$ is a parameter that captures the Van Hove singularity, $A(T)$ is the temperature-dependent exchange stiffness and $\omega_c\equiv\frac{A(T)}{4\hbar \epsilon}$ is a cutoff frequency. 
To avoid an explicit dependence on the experimental Curie temperature in these formulae, a Callen-Callen formula for the exchange stiffness \cite{callenPresentSatusTemperature1966} is adopted, where $A(T)=A_0 \langle S \rangle^2$, with $A_0=J_{ij}^{1NN}a^2$ being the exchange stiffness at $T=\qty{0}{\kelvin}$, $J_{ij}^{1NN}$ is the exchange constant up to the first nearest neighbors and $a$ is the atomic lattice constant.

\begin{table}[htbp]
    \centering
    \begin{adjustbox}{width=\columnwidth}
        \begin{tabular}{|c|c|c|c|c|c|}
        \hline
        \textbf{Element} & $J_{ij}^\text{MFA}$ & $\gamma_\text{SW}$ & $J_{ij}$ & $\epsilon$ & $\alpha$ \\
                        & (meV/link) & & (meV/link) & (\unit{\nano\metre\squared}) &  \\
        \hline
        Ni          & 17.2 \cite{evansAtomisticSpinModel2014}          & 0.79\cite{garaninSelfconsistentGaussianApproximation1996}       & 13.6  &$4\times 10^{-3}$ & 0.045 \cite{walowskiIntrinsicNonlocalGilbert2008} \\ 
        \hline
        Fe        & 38.2 \cite{wangExchangeInteractionFunction2010}          & 0.77\cite{garaninSelfconsistentGaussianApproximation1996}       & 30.2                                     & $2.85\times 10^{-3}$\cite{wooQuantumHeatBath2015} & 0.05 \cite{baratiGilbertDampingMagnetic2014} \\ 
        \hline
        Co            & 38.0 \cite{evansAtomisticSpinModel2014}          & 0.79\cite{garaninSelfconsistentGaussianApproximation1996}      & 30.0                                    & $4\times 10^{-3}$ & 0.015 \cite{baratiCalculationGilbertDamping2013} \\ 
        \hline
        \end{tabular}
    \end{adjustbox}
    \caption{Table of material parameters for the 3d ferromagnets.}
    \label{table:cp_elements}
\end{table}

By connecting the dLLB model to the quantum heat bath and incorporating the corrected exchange constant \(J_{ij}\), we achieve excellent agreement with experimental data for the three simulated ferromagnetic materials, as illustrated in Fig.~\ref{figure:magnetization-curve} (blue circles). 
This agreement demonstrates that incorporating the quantum heat bath into the dLLB model not only successfully captures the physics of magnons at low temperatures, but also improves the critical behavior near the Curie temperature. 

\begin{figure}[htbp]
    \centering
    \includegraphics[
    height=.8\textheight,
    width=\columnwidth,keepaspectratio]{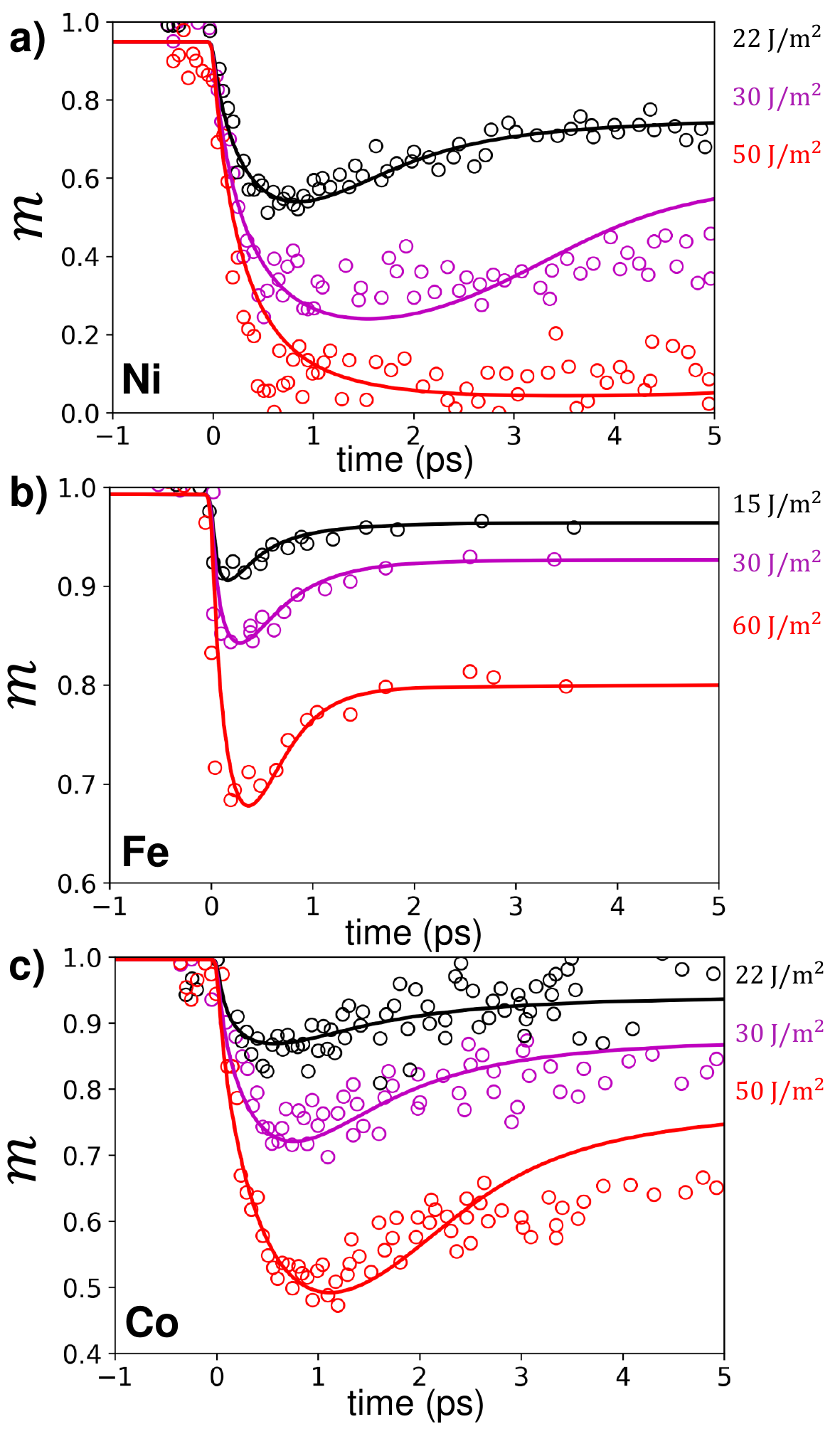}
    \caption{Norm of the volume average of the first moment as a function of time for a) Nickel, b) Iron, c) Cobalt, computed by atomistic spin dynamics of dLLB equations coupled to the 2TM with the parameters highlighted in Table.\ref{table:2TM_parameters}. The experimental data for Ni and Co are adapted from \cite{koopmansExplainingParadoxicalDiversity2010}, while for Fe, the data are adapted from \cite{carpeneDynamicsElectronmagnonInteraction2008}. The colors demonstrating an increase in experimental fluence are black, purple, and red, respectively.}
    \label{figure:ultrafast_demagnetization}
\end{figure}

%%%%%%%%%%%%%%%%%%%%%%%%%%%%%%%%%%%%%%%%%%%%%%%%%%%%%%%%%%%%%%%%%%
\subsection{Ultrafast demagnetization of 3d ferromagnets}\label{sec:LaserS}
%%%%%%%%%%%%%%%%%%%%%%%%%%%%%%%%%%%%%%%%%%%%%%%%%%%%%%%%%%%%%%%%%%

The dLLB model requires thermal connection to a spin reservoir, which can be provided by the 3TM. 
However, linking it to electron and phonon reservoirs requires effective coupling constants and heat capacities that are challenging to determine experimentally or compute using atomistic simulations.
Consequently, previous studies on atomistic ultrafast demagnetization either neglected such a reservoir or assumed that the phonon temperature should be considered because spins, as magnetic moments, are located on atoms. 
Unfortunately, at elevated fluences, the increase in the phonon temperature is not enough to replicate experimental data, whereas the electronic temperature appears to be more appropriate.
In the absence of suitable data, we simplify the 3TM to a 2TM by assuming that electron and spin temperatures are identical. 
Additionally, the coupling constant between electrons and phonons includes, to some degree, an effective reminder of the atomic magnetism.

We couple the dLLB-QFDR to a 2TM that describes the energy transfer between the electrons and phonons. 
In equations this can be expressed as
\begin{align}
    \begin{aligned}
    C_e\frac{dT_e}{dt}&=-G_{e-ph}(T_{ph}-T_e)+P(t)\\
    C_{ph}\frac{dT_{ph}}{dt}&=G_{e-ph}(T_{ph}-T_e)\\
    \textrm{with }P(t)&=\frac{\mu F}{\ell_{p}\tau}\exp(-\frac{(t-t_{0})^{2}}{\tau^{2}/4\operatorname{log}(2)})
    \end{aligned}
    \label{2TM}
\end{align}
where $C_e$ and $C_{ph}$ are the electron and phonon heat capacities respectively, $G_{e-ph}$ is the electron-phonon coupling constant, $\mu$ is the material absorbance, $F$ is the laser surface fluence, $\ell_{p}$ is the laser penetration depth, $t_0$ is the pulse delay and $\tau$ is the pulse duration. 
In Eqs.~\eqref{2TM}, the laser excitation has a Gaussian shape $P(t)$ in time. 
The thermal connection between such a 2TM to the dLLB model is performed by assuming that the electronic temperature is the source of magnetization fluctuation, entering the computation of $D$. 
The data are provided in Table~\ref{table:2TM_parameters}, with the approximation of the linear dependence of the electronic heat capacity with the electronic temperature $C_e=\gamma T_e$.
Both the 2TM and dLLB-QFDR dynamical models are solved simultaneously.

Results of ultrafast demagnetization for increasing laser fluences are shown in Fig. \ref{figure:ultrafast_demagnetization} (a), (b), and (c) for Ni, Fe, and Co, respectively.
The experimental demagnetization curves are provided by state-of-the-art references~\cite{koopmansExplainingParadoxicalDiversity2010, carpeneDynamicsElectronmagnonInteraction2008}.
The figures show good agreement between the dLLB-QFDR (solid line) model and the experimental data (symbols) for all the considered ferromagnetic elements. 
For nickel and cobalt, the model accurately captures the demagnetization rates and timescales across the explored fluence range in Fig.~\ref{figure:ultrafast_demagnetization}(a) and (b).
As the fluence increases, the transition from quick demagnetization followed by rapid re-magnetization to quick demagnetization with slower re-magnetization is accurately reproduced for Ni at high fluence. 
This behavior is obvious from the red curve ($F = \qty{50}{\joule\per\meter\squared}$) in Fig.~\ref{figure:ultrafast_demagnetization}(a), occurring when the electronic temperature significantly exceeds the Curie temperature of the ferromagnetic element. 
For iron, as shown in Fig.~\ref{figure:ultrafast_demagnetization}(b), a good agreement is again observed for both the demagnetization and re-magnetization regimes. 
However, achieving this agreement was only possible by considering the changes in electron heat capacity with increasing fluence. 
This result is expected because in general, it is moderately accurate to assume that electronic heat capacity depends linearly on temperature during laser excitation. 
Indeed, previous studies have shown that, for 3d metals at high temperatures, both $C_e$ and $G_{e-ph}$ exhibit a nonlinear dependence on temperature~\cite{heComputationalStudyShortPulse2019,linElectronphononCouplingElectron2008}.
The same approach is used for calculating the phonon heat capacity ($C_p$), which was believed to remain constant regardless the temperature, and equal to the phonon heat capacity at room temperature.
This approximation is known to be crude~\cite{srivastavaPhysicsPhonons2022}.
Although a more complex 2TM that naturally includes these effects might be needed to reproduce more accurately the experimental results, this approach is beyond the scope of the current work.

\begin{table}[htbp]
    \begin{adjustbox}{width=\columnwidth}
    \begin{tabular}{|c|c|c|c|c|c|c|c|}
        \hline
        \textbf{Element} & $C_p$             & $\gamma$           & $G_{e-ph}$       & $\mu$ & $\ell_p$ & $\tau$ & F\\
                         & (\unit{\joule\per\meter\cubed\per\kelvin}) & (\unit{\joule\per\meter\cubed\per\kelvin\squared}) & (\unit{\watt\per\meter\cubed\per\kelvin}) &       & (\unit{\nano\meter})       & (\unit{\femto\second}) & (\unit{\joule\per\meter\squared})\\
        \hline
        Ni               & $2.3\times10^6$ & $6\times10^3$       & $12\times10^{17}$ &0.5 &15 &50  & (22-50)\\ 
        \hline
        Fe    & $2.7\times10^6$  & $(2.5-5)\times10^3$ & $54\times10^{17}$ &0.3 &7 &60 &(15-60)\\ 
        \hline
        Co    & $2.1\times10^6$ & $3.1\times10^3$ & $12\times10^{17}$  & 1 & 15 &50 &(22-50)\\ 
        \hline
    \end{tabular}
\end{adjustbox}
\caption{Table of material parameters used for the 3d ferromagnetic single elements within the 2TM. The parameters are taken from \cite{koopmansExplainingParadoxicalDiversity2010, beaurepaireUltrafastSpinDynamics1996, medvedevElectronphononCouplingMetals2020}.  
}
\label{table:2TM_parameters}
\end{table}

In summary, the dLLB model effectively replicates ultrafast demagnetization experiments of 3d ferromagnetic elements. Compared to state-of-the-art simulations based on the stochastic Landau-Lifshitz-Gilbert (sLLG) equation, dLLB offers a faster and more accurate alternative due to its deterministic nature. This advantage arises because sLLG simulations require multiple stochastic realizations and averaging to produce reliable results. A direct comparison between dLLB and sLLG for the ultrafast demagnetization of Nickel is provided in Supplementary Note 1. 

\subsection{Ultrafast dynamic susceptibility of ferromagnets}\label{sec:LaserSigma}

To this point, we have focused on the laser-induced ultrafast dynamics of the first moment of the dLLB equations, specifically. 
In the ensuing phase, we aim to delve into the ultrafast dynamics of the second moments, as denoted by the tensor instantaneous volume average $\chi$, and defined in Eq.~\eqref{eq::second_moment}. 
The diagonal part of $\chi$ labelled $\chi_{II} = \left\langle S_I S_I \right\rangle - \left\langle S_I \right\rangle^2$ ($I=x,y,z$), serve to quantify the variance of spin orientation relative to their mean values.
Conversely, the off-diagonal part $\chi_{IJ} = \left\langle S_I S_J \right\rangle - \left\langle S_I \right\rangle \left\langle S_J \right\rangle$ ($I,J=x,y,z$, and $I\neq J$) represent the covariance, a measure of the joint variation of the spin orientations.

To gain a comprehensive understanding of the dynamics governing the $\chi$ quantity, we examine its behavior under specific conditions, focusing on Nickel. 
In this context, the magnetization is initially oriented along the $x$-axis, and the system is subjected to an irradiance of $F=\qty{40}{\joule\per\meter\squared}$. 

\begin{figure*}[htbp]
    \centering
    \includegraphics[
        width=\textwidth,keepaspectratio]{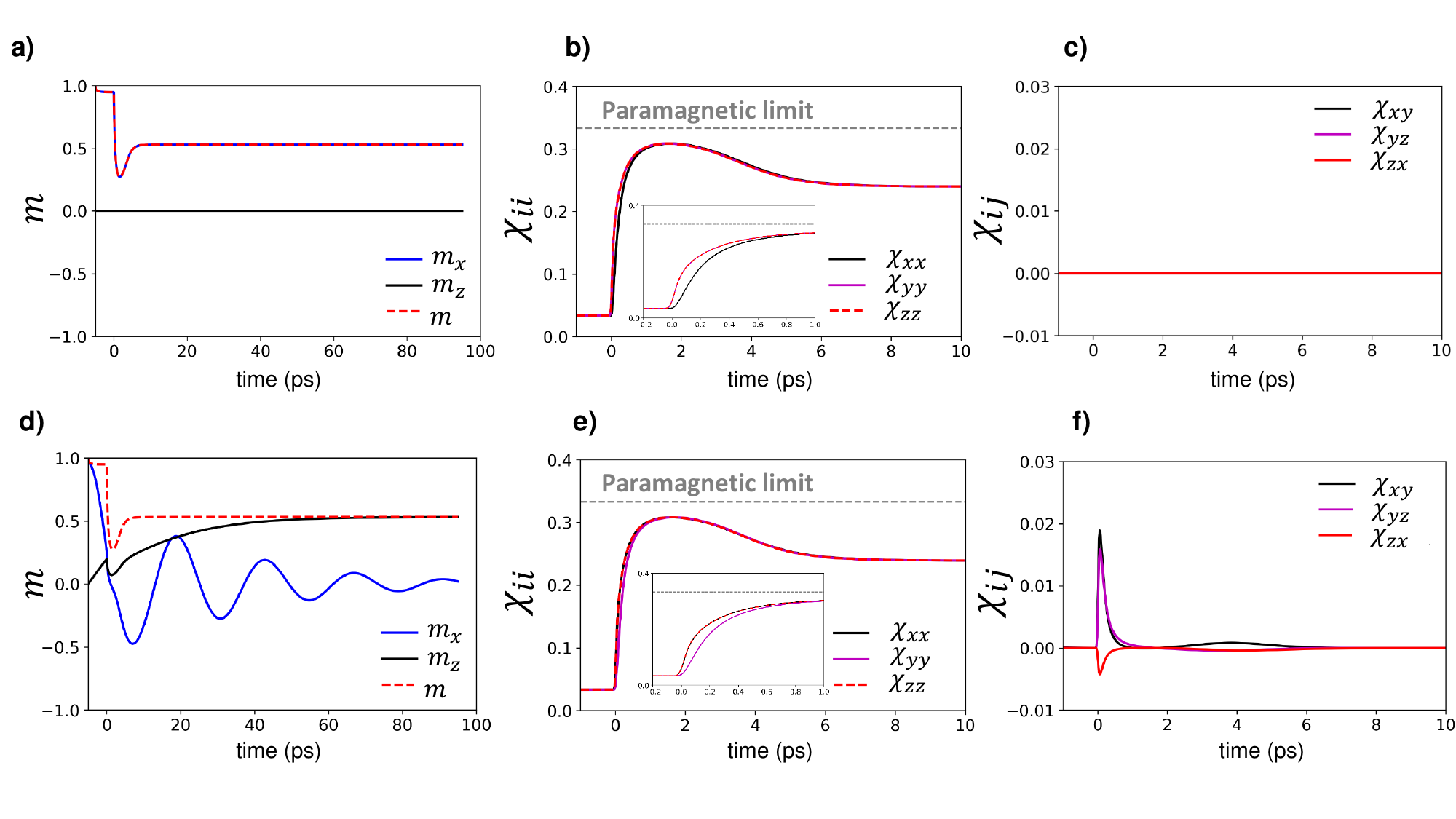}
    \caption{a) and d) Volume averaged first moment component as a function of time. b) and e) Diagonal elements of the second moment tensor of dLLB equation as a function of time. c) and f) Off-diagonal elements of second moment tensor of dLLB equation as a function of time. For a),b) and c) ${\bm B}={\bm 0}$. For d), e) and f) ${\bm B}=B_0{\bm e}_z$, with $B_0=1.5\,\mathrm{T}$ being the DC magnetic field.}
    \label{figure:UltrafastDemagMoments}
\end{figure*}
Two distinct scenarios are considered in our analysis.
First, we consider the isotropic situation, in which all spatial orientations have the same statistical weight. 
Subsequently, we examine a scenario in which an additional external DC magnetic field ${\bm B}$ is introduced along the $z$-axis.

In Fig.~\ref{figure:UltrafastDemagMoments}, the evolution of the average first moment components is presented along with the diagonal and off-diagonal parts of the second moment, $\chi$. 
In the absence of an external magnetic field, a typical ultrafast demagnetization process is observed in Fig.~\ref{figure:UltrafastDemagMoments}(a), similar to the processes discussed in Sec.~\ref{sec:LaserS}. 
On the other hand, Fig.~\ref{figure:UltrafastDemagMoments}(b) shows that $\chi_{xx}$, $\chi_{yy}$, and $\chi_{zz}$ initially increase, reaching a maximum around $\sim 0.3$, before decreasing and stabilizing around $\sim 0.25$. 
This behavior can be intuitively understood by considering that the laser pulse raises the temperature, which increases fluctuations in the spin system, ultimately leading to higher values of the variance $\chi_{ii}$. 
Interestingly, as the spin system enters the paramagnetic phase, the value of $\chi_{ii}$ approaches $\frac{1}{3}$, revealing that the 3 directions of space are equivalent, as expected by isotropy.
In Fig.~\ref{figure:UltrafastDemagMoments}(c), the off-diagonal elements of $\chi$ are plotted as a function of time. 
It is observed that $\chi_{xy}$, $\chi_{yz}$, and $\chi_{zx}$ remain zero over time. 
This behavior is due to the lack of correlation between the spatial orientations of the spin, once again due to the isotropic nature of the system.

Now, if an external DC magnetic field ${\bm B}$ is applied along the $z$-direction, the isotropy is broken and the overall behavior changes, as shown in Fig.~\ref{figure:UltrafastDemagMoments}(d), (e), and (f). 
From Fig.~\ref{figure:UltrafastDemagMoments}(d), it is observed that the ultrafast demagnetization process is preserved, but, because of the presence of the DC field, the magnetization undergoes Larmor precession before settling into a state where each spin is perfectly aligned with its local field. 
Interestingly, as demonstrated in Fig.~\ref{figure:UltrafastDemagMoments}(e), the components $\chi_{xx}$, $\chi_{yy}$, and $\chi_{zz}$ mirror the behavior observed when the external field is absent.
The off-diagonal elements of $\chi$, shown in Fig.~\ref{figure:UltrafastDemagMoments}(f), display a peak when the laser pulse is applied, followed by a bump before relaxing to zero.
Such a behavior can be ascribed to the external magnetic field disrupting the statistical equivalence between the various spatial orientations of the spins, consequently leading to the development of correlations among them.
This correlation becomes evident when we examine $\chi_{zx}$, which represents the covariance between $S_x$ and $S_z$. 
Under the influence of ${\bm B}$, $S_z$ is expected to increase while $S_x$ decreases. 
Consequently, the covariance between these two variables should be negative, as clearly illustrated by the red curve in Fig.~\ref{figure:UltrafastDemagMoments}(f).

Physically, the tensor $\chi(t)$ represents a dynamic susceptibility. 
The expression derived in this work is mathematically similar to that presented by Kubo {\it et al.}~\cite{kuboFluctuationdissipationTheorem1966} as it quantifies the fluctuation-dissipation relationship. 
The key difference lies in the treatment of spatial correlations: while the expression reported by Kubo accounts for spatial correlations, our formulation neglects them due to the mean-field approximation (MFA) of the exchange interaction.
This distinction highlights that, although the two susceptibilities share a similar mathematical structure, they are fundamentally distinct in their interpretations. 
Specifically, the susceptibility presented in~\cite{kuboFluctuationdissipationTheorem1966} quantifies fluctuations between neighboring spins within a ferromagnet close to the equilibrium, whereas the susceptibility in this study characterizes fluctuations in the spatial orientation of each individual spin in the transient regime, whether evaluated locally on each atomic site or globally when the volume average is computed.

%%%%%%%%%%%%%%%%%%%%%%%%%%%%%%%%%%%%%%%%%%%%%%%%%%%%%%%%%%
\section{Conclusion}\label{sec:conclusion}
%%%%%%%%%%%%%%%%%%%%%%%%%%%%%%%%%%%%%%%%%%%%%%%%%%%%%%%%%%

We have demonstrated that the dLLB method is an effective tool for modeling thermal effects in magnetic materials. 
Combined with a quantum fluctuation-dissipation theorem, this provides an excellent prediction of the Curie temperature as well as the magnetization curve at low temperatures, capturing the main features of the equilibrium state. 
A key advantage of the dLLB lies in its predictive nature, particularly compared to the classical LLB equation, which requires prior knowledge of precise thermodynamic functions, up to the Curie temperature of the simulated system.
For ultrafast dynamics, the dLLB equations demonstrated excellent agreement with experimental data for 3d ferromagnetic elements, effectively capturing the principal characteristics of the transient dynamics.
This accuracy in predicting the underlying physics is complemented by another significant advantage: numerical efficiency. 
The deterministic nature of the dLLB results offers a clear benefit for atomistic simulations. 
Compared to contemporary simulations utilizing the sLLG equation, the proposed dLLB method demonstrates a remarkable increase in computation speed, more than a hundred time faster in our case, with the same physical outcomes. 
The observed speedup is attributed to the need of performing numerous stochastic realizations in sLLG simulations to achieve reliable averaged values, critical for making definitive statements regarding the underlying physical system.
Furthermore, the dLLB approach is not confined to ferromagnets and can be expanded effortlessly to ferrimagnets and antiferromagnets. 
As an inherently atomistic method, it can also be tailored for examining multilayers with various types of magnetic interactions.

\section*{Acknowledgments}\label{sec:acknowledgments}
This work is partially supported by the France 2030 government investment plan managed by the French National Research Agency under grant reference PEPR SPIN – [SPINTHEORY] ANR-22-EXSP-0009, and CEA Exploratory Program, Bottom-Up 2023 ref.59 [THERMOSPIN].

%\bibliographystyle{apsrev4-2}
%\bibliography{paper}

\end{document}

\def\mytitle{Supplementary Materials:\\ Ultrafast dynamics of moments in bulk ferromagnets}
\def\myfirstauthor{Mouad Fattouhi}
\def\mysecondauthor{Pascal Thibaudeau}
\def\mythirdauthor{Liliana D. Buda-Prejbeanu}
\def\spintec{Univ. Grenoble Alpes, CNRS, CEA, SPINTEC, F-38000, Grenoble, France}
\def\dam{CEA, DAM, Le Ripault, F-37260, Monts, France}

\title{\mytitle}
\author{\myfirstauthor}
\email{mouad.fattouhi@cea.fr}
\affiliation{\spintec}
\author{\mysecondauthor}
\affiliation{\dam}
\author{\mythirdauthor}
\affiliation{\spintec}

\maketitle

\onecolumngrid 

\section{Ultrafast demagnetization of Nickel: Dynamic Landau\--Lifshitz\--Bloch vs stochastic Landau-Lifshitz-Gilbert equation}\label{sup:UltFast-Ni}

\begin{figure}[htbp]
    \centering
    \sidesubfloat[]{\includegraphics[width=0.45\textwidth]{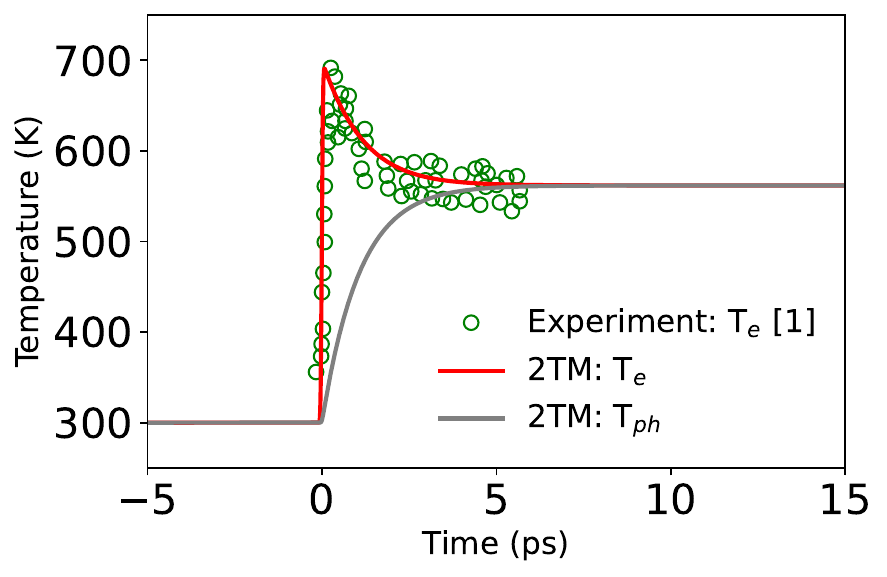}}
    \\
    \sidesubfloat[]{\includegraphics[width=0.45\textwidth]{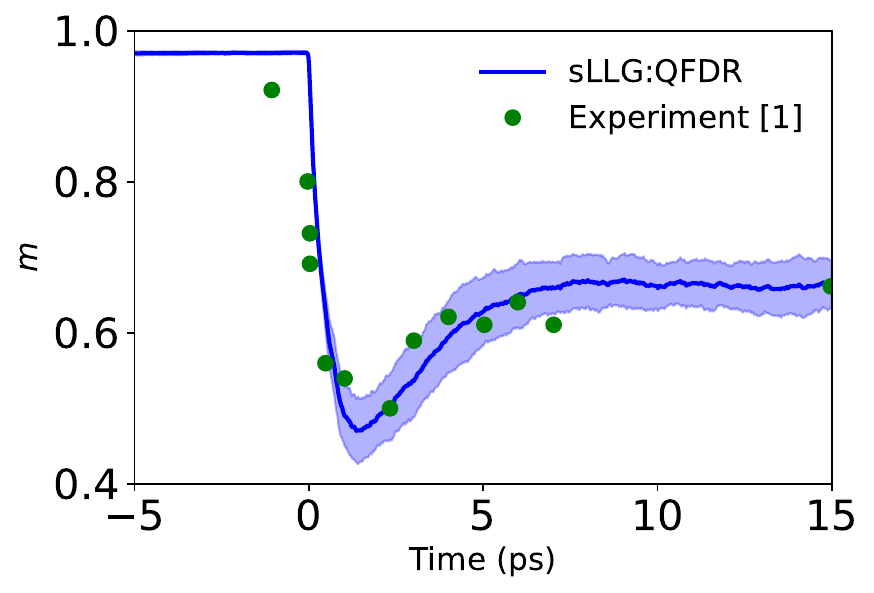}}
    \hfill
    \sidesubfloat[]{\includegraphics[width=0.45\textwidth]{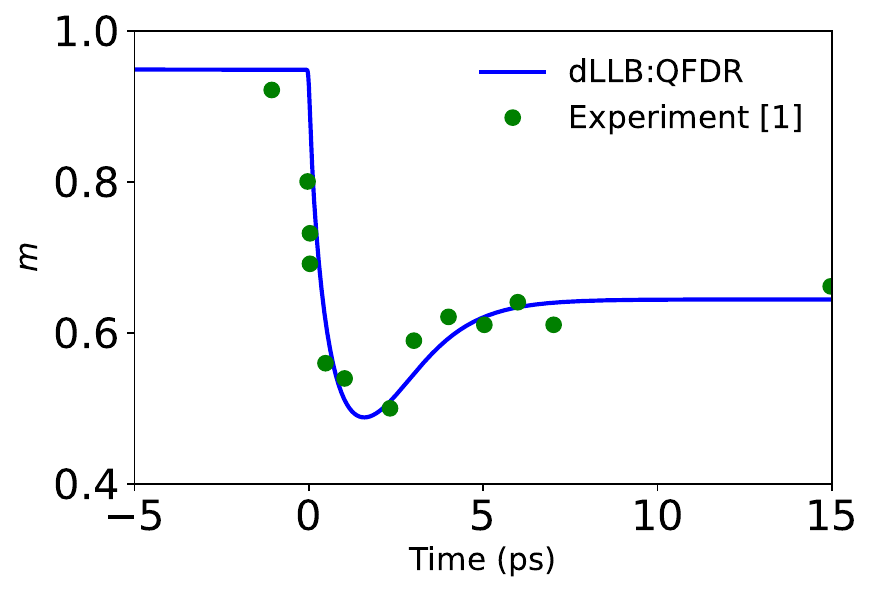}}
    \caption{a) Electron and phonon temperature computed from the numerical solution of the two-temperatures model and compared to the experimental electron temperature of Nickel, taken from \cite{beaurepaireUltrafastSpinDynamics1996}. b) Volume average of the first moments versus time, as computed from more than one hundred realizations based on the atomistic sLLG equation with QFDR, under the experimental condition of ref.\cite{beaurepaireUltrafastSpinDynamics1996}; The simulated system consisted of 500 atoms with periodic boundary conditions. c) Volume average of the first moments versus time as computed from one single integration dLLB equation with QFDR.}
\label{Sfig:2TM}
\end{figure}

In this section, we present a comparison between the dynamic Landau-Lifshitz-Bloch (dLLB) equations and the stochastic Landau-Lifshitz-Gilbert (sLLG) equation in predicting the ultrafast demagnetization dynamics of Nickel, as experimentally observed in \cite{beaurepaireUltrafastSpinDynamics1996}.
To achieve this, we employ an in-house atomistic spin dynamics code, which implements the atomistic sLLG formalism~\cite{thibaudeauThermostattingAtomicSpin2011}. 
The sLLG equation is coupled along with a quantum fluctuation-dissipation relationship (see main text) and a two-temperature model, that takes into account laser-induced heating.
All material parameters used in these simulations are identical to those used in the main manuscript for Nickel.
Figure~\ref{Sfig:2TM}(a) displays both the electron and phonon temperature evolution as computed from the numerical solution of the two-temperature model (2TM), alongside the reported experimental data from~\cite{beaurepaireUltrafastSpinDynamics1996}.
The material parameters used to solve the 2TM were obtained from Table II of the primary paper.
The results demonstrate excellent agreement between the computed and experimental electronic temperature curves.
On one hand, the blue plain line of figure~\ref{Sfig:2TM}(b) illustrates the dynamics of the volume average of first moments, obtained from the numerical solution of the coupled sLLG equations, averaged over more than one hundred of stochastic realizations of the noisy dynamics. 
This graph also features a shaded area, which computes the standard deviation of these stochastic simulations, serving to determine the confidence interval of these realizations.
On the other hand, the figure~\ref{Sfig:2TM}(c) presents the dynamics of the volume average of the first moments as a function of time, obtained from a single numerical solution of the dLLB equations. 
This result highlights the efficiency of the dLLB equation in accurately predicting the magnetization dynamics of nanomagnetic systems, within the transient regime, all while offering significantly faster computational performance as opposed to numerous simulations of the sLLG equation.

%*
%\begin{figure}
    %\begin{center}
    %\includegraphics[width=\textwidth]{Fig2S}
    %\begin{minipage}[b]{0.5\linewidth}
      %  \centering
     %   \sidesubfloat[]{\includegraphics[width=0.9\textwidth,keepaspectratio=true]{2TM.pdf}}
    %\end{minipage}
    %\begin{minipage}[t]{0.49\linewidth}
        %\centering
   %     \sidesubfloat[]{\includegraphics[width=0.9\textwidth,keepaspectratio=true]{sLLG.pdf}}\par\vspace{5pt}
  %      \sidesubfloat[]{\includegraphics[width=0.9\textwidth,keepaspectratio=true]{dLLB.pdf}}
 %   \end{minipage}
%\end{center}
%\caption{a) Electron and phonon temperature computed from the numerical solution of the two-temperatures model and compared to the experimental electron temperature of Nickel, taken from \cite{beaurepaireUltrafastSpinDynamics1996}. b) Volume average of the first moments versus time, as computed from more than one hundred realizations based on the atomistic sLLG equation with QFDR, under the experimental condition of ref.\cite{beaurepaireUltrafastSpinDynamics1996}; The simulated system consisted of 500 atoms with periodic boundary condition. c) Volume average of the first moments versus time as computed from one single integration dLLB equation with QFDR.}
%\label{Sfig:2TM}
%\end{figure}

\section{Ultrafast dynamics of the second moment under different laser pulse durations and fixed fluence}\label{sup:introduction}

\begin{figure}[htbp]
    \centering
    \sidesubfloat[]{\includegraphics[width=0.45\textwidth]{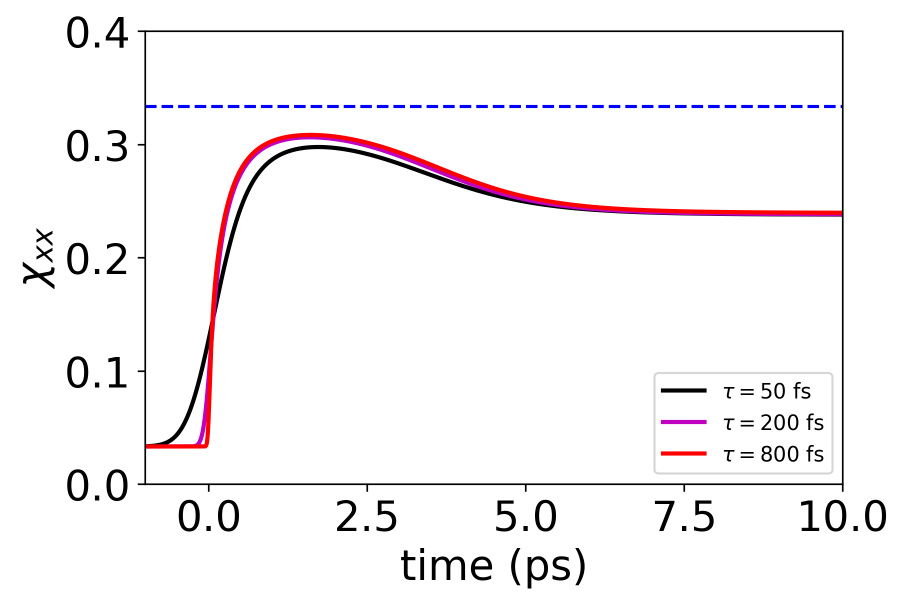}}
    \hfill
    \sidesubfloat[]{\includegraphics[width=0.45\textwidth,height=52mm]{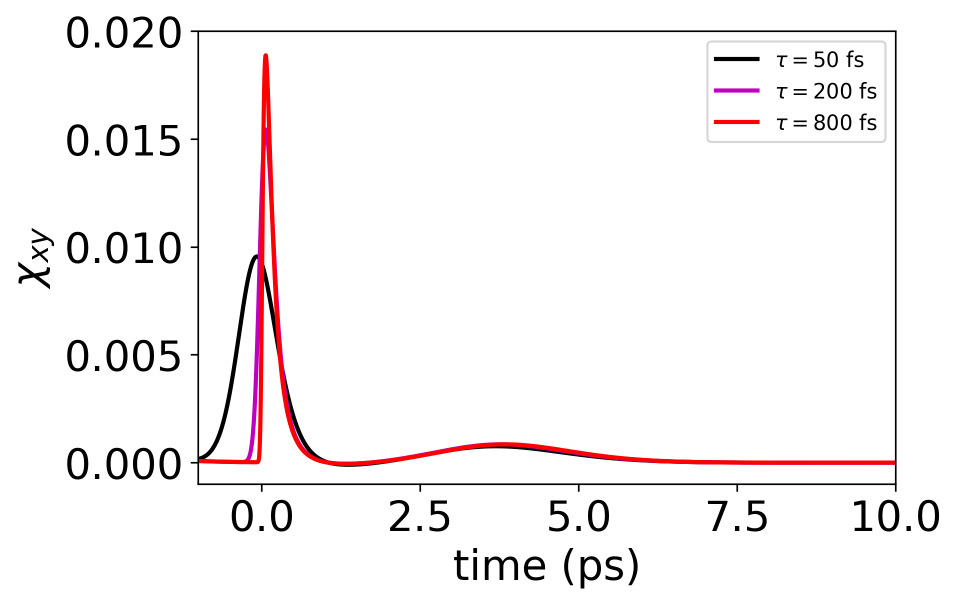}}
    \\
    \sidesubfloat[]{\includegraphics[width=0.45\textwidth]{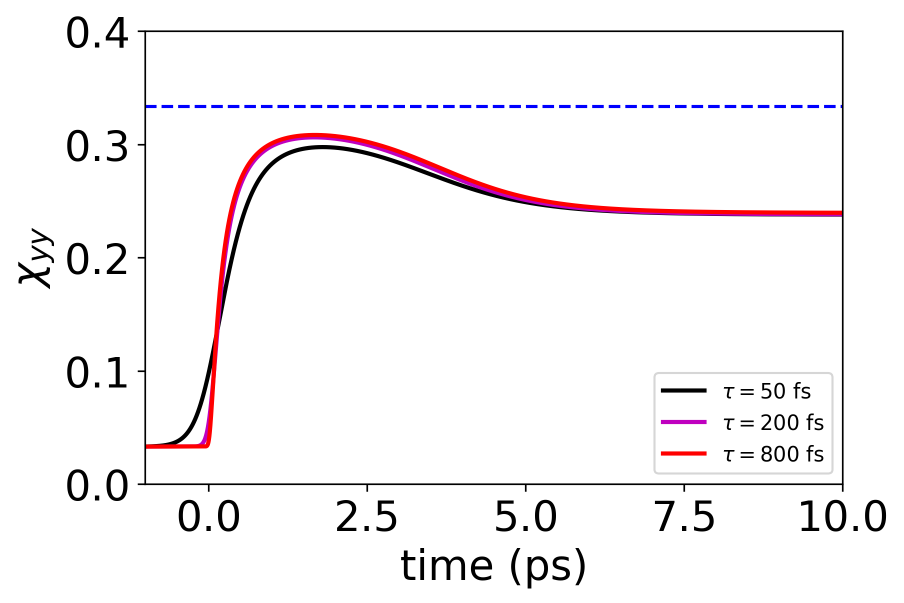}}
    \hfill
    \sidesubfloat[]{\includegraphics[width=0.45\textwidth,height=52mm]{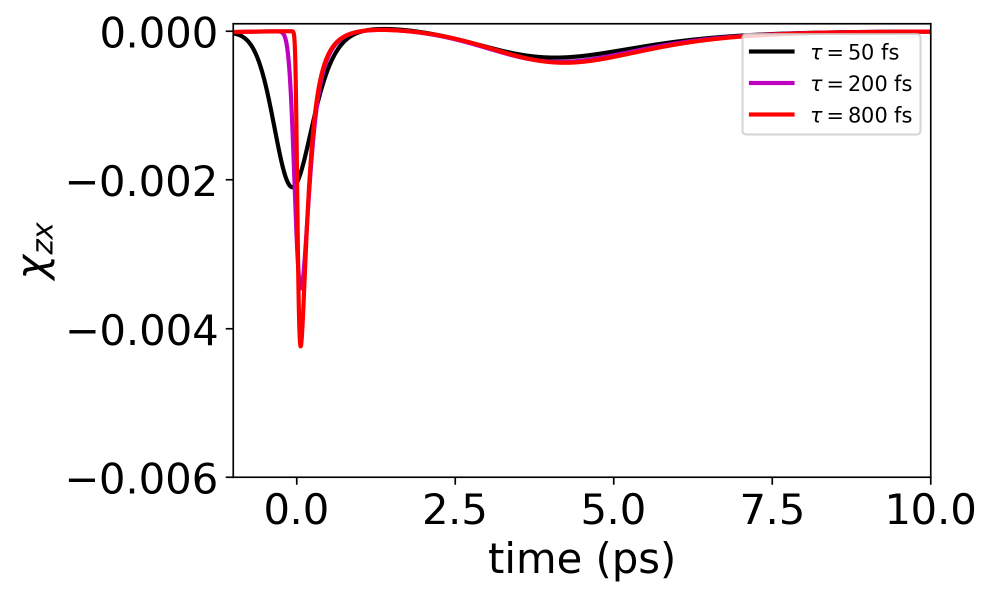}}
    \\
    \sidesubfloat[]{\includegraphics[width=0.45\textwidth]{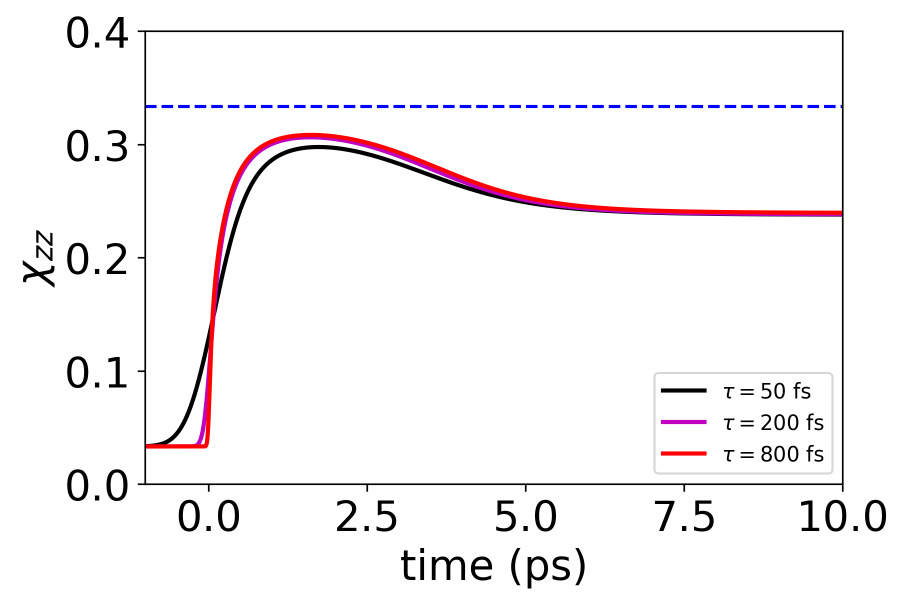}}
    \hfill
    \sidesubfloat[]{\includegraphics[width=0.45\textwidth,height=52mm]{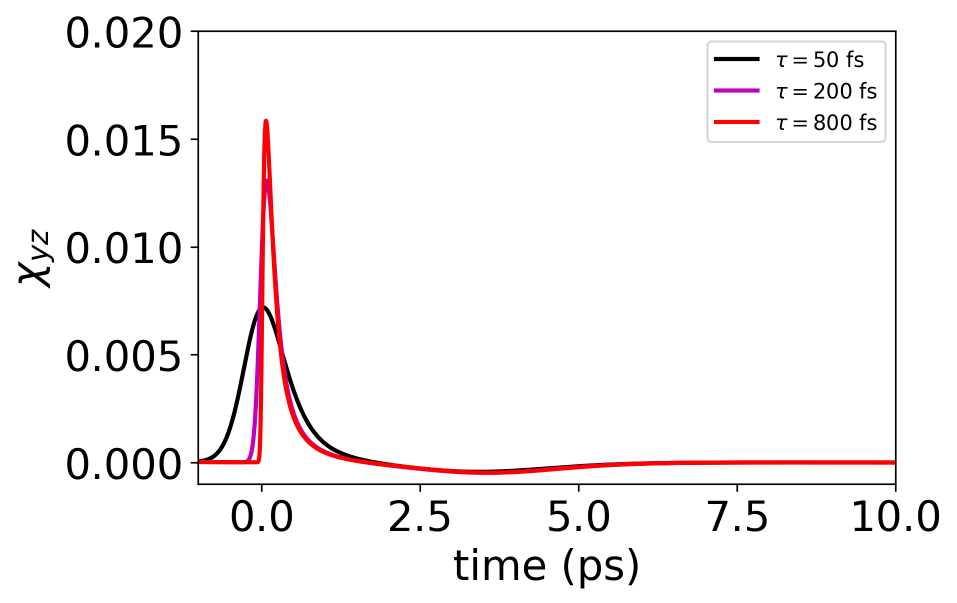}}
    \caption{Components of the second cumulant ${\chi}$ for three pulse durations and constant external magnetic field ${\bm B}=B_0{\bm e}_z$ with $B_0=1.5\,\mathrm{T}$, a) $\chi_{xx}(t)$. b) $\chi_{xy}(t)$. c) $\chi_{yy}(t)$. d) $\chi_{zx}(t)$. e) $\chi_{zz}(t)$. f) $\chi_{yz}(t)$.}
    \label{Sfig:secondmoments}
\end{figure}

In this section, we present additional results regarding the ultrafast dynamics of a laser pulse with varying pulse durations for a fixed fluence and a constant DC magnetic external induction along the $z$-axis. 
Simulations have been conducted for laser pulse widths, which include $\tau=\qty{50}{\fs}$, $\qty{200}{\fs}$, and $\qty{800}{\fs}$, while maintaining a fluence of $F=\qty{40}{\joule\per\metre\squared}$ and a constant magnetic field strength of $B_0=\qty{1.5}{\tesla}$.
For $t=0$, the magnetization is aligned along  
The figure~\ref{Sfig:secondmoments} shows the evolution over time of the diagonal and off-diagonal parts of the tensor $\chi$ for three different pulse durations.
The definition of $\chi$ is given in the main document. 
As observed in the figure, the three diagonal components, $\chi_{xx}$, $\chi_{yy}$, and $\chi_{zz}$, exhibit minimal sensitivity to changes in pulse duration, that is consistent to the isotropy condition.
In contrast, the off-diagonal parts $\chi_{xy}$, $\chi_{yz}$, and $\chi_{zx}$ display significant variations as the pulse duration increases. 
Specifically, as the pulse duration lengthens, the peak amplitude decreases while its width broadens.
Furthermore, the bump observed after the laser pulse has vanished remains practically unchanged across the three cases shown in Fig.~\ref{Sfig:secondmoments}(d), (e), and (f).

%\bibliographystyle{apsrev4-2}
%\bibliography{spintronics}
